\documentclass[conference]{IEEEtran}
\IEEEoverridecommandlockouts
\usepackage{cite}
\usepackage{amsmath,amssymb,amsfonts}
\usepackage{algorithmic}
\usepackage{graphicx}
\usepackage{textcomp}
\usepackage{xcolor}
\usepackage{underscore}
\usepackage{tabularx}
\usepackage{comment}

\newcounter{magicrownumbers}

\newcommand\vS{Spoofing}
\newcommand\vT{Tampering}
\newcommand\vR{Repudiation}
\newcommand\vI{Information Disclosure}
\newcommand\vD{Denial of Service}
\newcommand\vE{Elevation of Privilege}

\def\BibTeX{{\rm B\kern-.05em{\sc i\kern-.025em b}\kern-.08em
    T\kern-.1667em\lower.7ex\hbox{E}\kern-.125emX}}

\makeatletter

\def\ps@IEEEtitlepagestyle{
  \def\@oddfoot{\mycopyrightnotice}
  \def\@evenfoot{}
}
\def\mycopyrightnotice{
  \gdef\mycopyrightnotice{}
}

\@ifundefined{showcaptionsetup}{}{
 \PassOptionsToPackage{caption=false}{subfig}}
\usepackage{subfig}
\makeatother

\usepackage{eso-pic}
\newcommand\AtPageUpperMyright[1]{\AtPageUpperLeft{
 \put(\LenToUnit{0.5\paperwidth},\LenToUnit{-1cm}){
     \parbox{0.5\textwidth}{\raggedleft\fontsize{9}{11}\selectfont #1}}
 }}
\newcommand{\conf}[1]{
\AddToShipoutPictureBG*{
\AtPageUpperMyright{#1}
}
}

\conf{Accepted in \emph{2021 3rd International Conference on Sustainable Technologies for Industry 4.0 (STI), 18-19 December, Dhaka}}
	    
\begin{document}
	
	\title{STRIDE-based Cyber Security Threat Modeling for IoT-enabled Precision Agriculture Systems}
	
	\author{\IEEEauthorblockN{Md. Rashid Al Asif\IEEEauthorrefmark{1}, Khondokar Fida Hasan\IEEEauthorrefmark{2}, Md Zahidul Islam\IEEEauthorrefmark{3}, and Rahamatullah Khondoker\IEEEauthorrefmark{4}}
	\IEEEauthorblockA{Department of Computer Science and Engineering, University of Barishal, Bangladesh\IEEEauthorrefmark{1}\\
	Centre for Cyber Security Research \& Innovation, RMIT University, 124 La Trobe Street, Melbourne, 3000, VIC, Australia\IEEEauthorrefmark{2}\\
	Department of Information and Communication Technology, Islamic University, Bangladesh\IEEEauthorrefmark{3}\\
	Department of Business Computing, THM University of Applied Sciences, Friedberg, Germany\IEEEauthorrefmark{4}\\
	Email: rashid.al.asif\IEEEauthorrefmark{1}@gmail.com,  fida.hasan\IEEEauthorrefmark{2}@rmit.edu.au,  zahidimage\IEEEauthorrefmark{3}@gmail.com,
	rahamatullah.khondoker\IEEEauthorrefmark{4}@mnd.thm.de}
	}
	\maketitle

\begin{abstract}
The concept of traditional farming is changing rapidly with the introduction of smart technologies like the Internet of Things (IoT). Under the concept of smart agriculture, precision agriculture is gaining popularity to enable Decision Support System (DSS)-based farming management that utilizes widespread IoT sensors and wireless connectivity to enable automated detection and optimization of resources. 
Undoubtedly the success of the system would be impacted on crop productivity, where failure would impact severely. Like many other cyber-physical systems, one of the growing challenges to avoid system adversity is to ensure the system’s security, privacy, and trust. But what are the vulnerabilities, threats, and security issues we should consider while deploying precision agriculture? This paper has conducted a holistic threat modeling on component levels of precision agriculture's standard infrastructure using popular threat intelligence tools STRIDE to identify common security issues. Our modeling identifies a noticing of fifty-eight potential security threats to consider. This presentation systematically presented them and advised general mitigation suggestions to support cyber security in precision agriculture.

\end{abstract}
	\section{Introduction}
The adoption and acceleration of disruptive technologies such as the Internet of Things (IoT) and cloud computing are propelling the 4th industrial revolution, changing our lives and environment remarkably. One of the smart moves of these technologies is in the farming system that replaces the traditional farming infrastructure with smart infrastructure. Precision agriculture is a smart farming management system that evolves from IoT aggregated sensors, actuators, and devices with the intention of interaction, control, and automated decision-making. Its already been introduced in many countries and rolling in some developing countries like Bangladesh and India to facilitate with lessen human effort, reduced cost, saving time while increasing harvest and profit \cite{b1}.

Most IoT devices that are being used for field sensings, such as temperature, humidity, and moisture sensors, are generally resource-constrained — they usually don’t have high-end computation capabilities or memory. The communication technology that has been employed also does not offer large bandwidth.  Such resource-constrained infrastructures are mal-reputed with system vulnerabilities and windows with attack vectors. Additionally, with the recent trend of cloud-based virtualization, Machine learning and Artificial Intelligence stepped forward the automation further by assisting with any intelligent and automated decision decision including remotely turning on/off an irrigation actuator, productivity monitoring and forecasting, and disease control \cite{b2}. There are many IoT platforms (both commercial and open-source) available for precision agriculture services in the marketplace \cite{b3}. However, involving the Internet undoubtedly widening the attack surface where attack vectors originate from global sites too \cite{b4}. Therefore, cyber threats in smart farming management such as precision agriculture are a significant concern for sustainable development that can directly impact crop growth and farmers' realization. This paper has identified the cyber threat associated with standard precision agriculture infrastructure to offer a mitigation strategy. 

Threat modeling is a well-accepted measure to design cyber-secure infrastructure that helps identify, enumerate, prioritize threats for a system. It is a part of threat intelligence aimed to apply appropriate controls against threats. There are different threat modeling tools available; however, we adopted STRIDE, a Microsoft corporation product, in our modeling. While developing our model, we consider the design at the component level to list all possible attack vectors within and between nodes. Our proposed model identifies a noticeable fifty-eight cyber security threats in a standard precision agriculture system. Considering these threats can assist in developing appropriate countermeasures towards a cyber-secure robust system.

The rest of this paper is structured as follows. Section II reviews the usage benefit of IoT and research scope toward security issues present at the device/component level in precision agriculture. Afterward, the research methodology is presented in Section III. Section IV summarizes components, data flows, interactors, and threat models to analyze threats. Section V lists identified cyber security threats plus a recommended list of defense mechanisms against threat categories. The paper is concluded in Section VI with the future scope.

	\section{Literature Review}
	Smart farming or precision agriculture enables farmers to treat plants (or animals) precisely with their needs towards highest level of productivity. Smart farming refers to the introduction of a range of advanced technologies including robotics, sensor technology, satellite imaging, GPS, image processing, big data, artificial intelligence, cloud computing. Besides, there are IoT platforms available to aid precision agriculture, such as Arable Mark (Pulsepod), Agri M2M, Libelium–Waspmote, A3-uRADMonitor, Observant. I. Marcu et al. presents a good comparative study on the basis of the features, disadvantages of those platforms \cite{b3}.
	
	A. R. de Araujo Zanella et al. reviewed security issues and their current state on smart agriculture \cite{b5}. They identified smart agriculture as still in the emerging stage with low-level security features (still an open issue). But, building efficient and robust system security is crucial. Later on, four-layered smart-agricultural elements are presented along with possible security issues. Finally, they provide a list of security resources (i.e., intrusion/anomaly detection system, firewall, anti-malware, anti-virus, access control, authentication, cryptography) where further improvement may enhance the security in smart agriculture as a future direction.
	
	Another study carried on security issues covering precision agriculture and introduced a risk-based framework using Common Vulnerability Scoring System (CVSS) as a basis to assess and prioritize vulnerabilities \cite{b6}. The proposed framework helps to understand the cyber security vulnerabilities within the technology itself and the environment wherein it is adapted. Therefore, it helps to determine the required elements for constructing a precision agriculture system protected from cyber-attacks.
	
	R. Khan et al. presented a simplified threat modeling framework for cyber-physical systems (using Microsoft STRIDE) to ensure system security at the component level \cite{b7}. But it is a prerequisite to address vulnerabilities for each component whenever interdependencies between components. Otherwise, the entire security system may fall out of control. Therefore, the framework followed a systematic and comprehensive approach to maximize security at the component level.  So, it might be helpful for a system design and validation process before deployment.
	
	A. Omotosho et al. performed threat modeling on eleven IoT health devices based on device assets and access points \cite{b8}. They have employed the STRIDE model to identify device threats and rank them using a threat-risk model named DREAD. Also, they suggested countermeasures to mitigate identified threats. Moreover, as part of the model prototype, they developed a web-based system where all stakeholders see device vulnerabilities, privacy risks, and threat ratings. Thus, create an opportunity for designers to improve product security. M. Cagnazzo  et al. represented a similar threat modeling in the context of the Mobile Health (mHealth) System \cite{b9}. But it was with a focus on encryption and authentication for resource-constrained devices. From a mHealth prototype, authors identify assets and classify threats using STDIRE methodology. Like the previous work, it labeled risk-level using DREAD and recommend possible mitigation strategies to afford a reliable atmosphere.
	
	Refer to above works, threat modeling resultant details may be helpful in mitigation strategies to increase system security. Moreover, researchers from \cite{b10} presented threat modeling on a generic telesurgery system considering its components and data flows. The resultant threat detail might be crucial to check whether proper mitigations are present before deployment.
	
	In this work, we consider a precision agriculture scenario based on the proposed IoT architecture \cite{b11} to perform threat modeling as an extension of previous work.
	
	\section{Methodology}
\begin{figure}
  	\centerline {\includegraphics[width=3.3in]{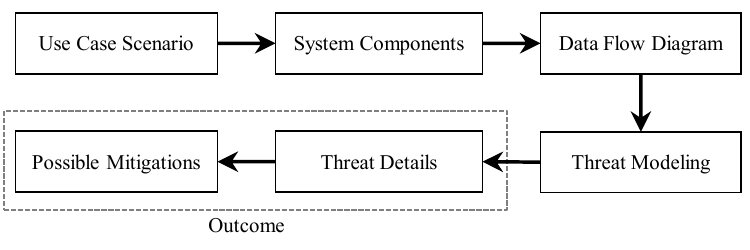}}
 	\caption{Consolidated Scheme of Threat Modeling}
 	\label{fig-steps}
	\end{figure}

	This research aims to perform threat modeling of precision agriculture by following the approach displayed in Fig. \ref{fig-steps}. In its process, we have considered and added two use cases following one of our previous work\cite{b11}. This leads us to derived system components that reveal functionalities, relationships, and interactions among them. Next, we included a  data flow diagram to point all components and data flows. Afterward, the Microsoft STRIDE method is applied to perform threat modeling on the data flow diagram. As a result, the threat model produced threats with further details. Finally, we presented possible cyber security defense mechanisms to aid the mitigation strategies against identified threats.
	
 \begin{figure}[!h]
    \centerline{\includegraphics[width=2.50in, height=2.3in]{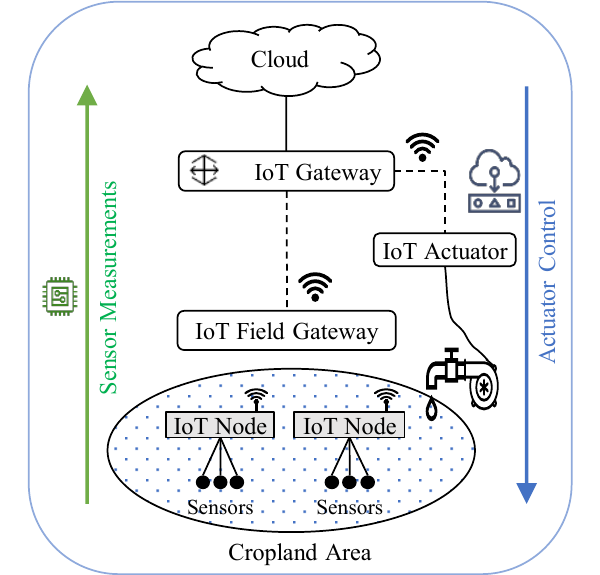}}
    \caption{Use cases of sensor measurement and actuator control}
    \label{fig-usecase}
    \end{figure}

	\subsection{Use Case Scenario}
	The use case intends to explain precision agriculture from the perspective of the proposed IoT architecture \cite{b11}. Simply, sensors are sensing measurements data (i.e., temperature, humidity, soil nutrient levels, pH, relative humidity, leaf-wetness) and sending to the cloud for further processing. Besides, remote users perform any actions (i.e., spraying, irrigation) via the actuators to the cropland.  From Fig. \ref{fig-usecase}, sensor measurements are sent from the cropland to the cloud. Besides, actuator control instructions are coming from the cloud to the actuator. Here, we have explained two use cases whether other use cases are possible to incorporate within the system.

	\paragraph{Sensing measurements}
	Cropland sensors are located on different locations of the imagined cropland. All of those are connected to the gateway and capable of sensing measurements frequently or on user demands. The role of the gateway is to collect and transmit those measurements to the cloud. As measurement data is stored in the cloud the remote users/cropland owners utilize those for purposes.
	\paragraph{Device control}
	If the sensed soil humidity is below the standard range then it should start the irrigation process. In that case, the IoT ecosystem sends start instruction to the Salo actuator. Besides, the cropland sensors are sending live humidity measurements frequently to the cloud. At the same time, the IoT system continuously checks certain humidity levels whether to stop the irrigation process. And this can be done automatically by setting appropriate conditions for specific measurements and actions for actuators.

\begin{figure}[!h]
  	\centerline {\includegraphics[width=.48\textwidth]{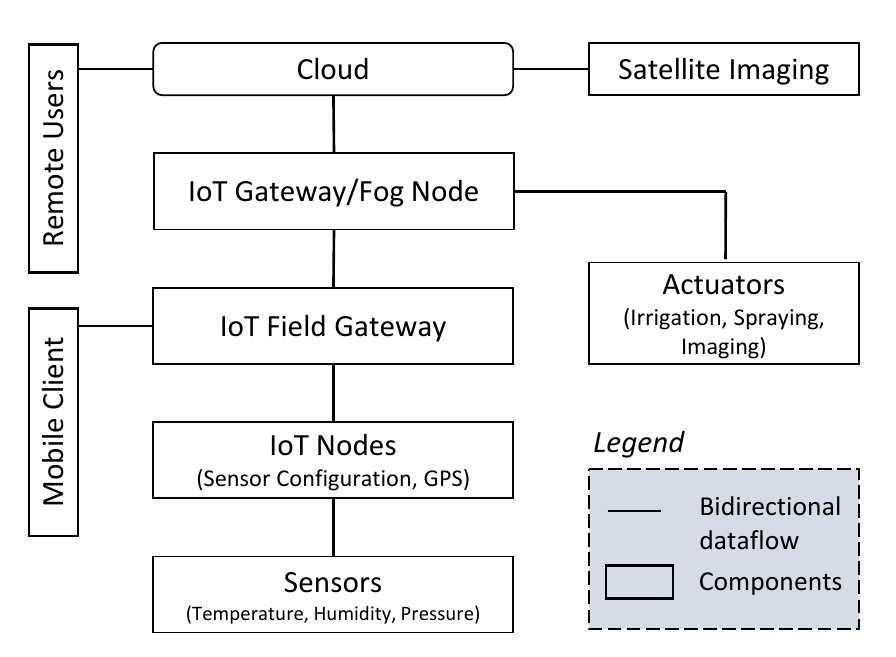}}
 	\caption{System components of precision agriculture}
 	\label{fig-blockdiagram}
	\end{figure}
	

	\subsection{System Components}
	Fig. \ref{fig-blockdiagram} illustrated the system components of precision agriculture, derived to specify the function of components, relationship, and interaction among them. Each component performs some actions based on input and generates output. The generated output may feed into the input of another component. Moreover, one component may have multiple input dependencies to make it work. As a whole, the components including gateways, sensors, cloud services, IoT nodes, and human interactors have their role and integrated part of the precision agriculture system.
	
    With technological advancement (i.e., IoT, robotics, artificial intelligence, sensor technology, satellite imaging, GPS, image processing, etc.), precision agriculture allows better crops production by reducing human effort \cite{b12}. It enables farmers to monitor crop health for large areas in real-time. However, it is essential to capture croplands or sense several parameters such as soil temperature, pressure, humidity, soil nutrient levels, pH, relative humidity, leaf wetness, and others. Generally, sensors sense various parameters to send the cloud via edge devices. Consequently, those measurements are analyzed to take further actions like spraying, irrigation. Finally, the actuator performs the action based on the analyzed decision. Moreover, several issues (i.e., soil fertilization, disease forecasting, and detection) can be addressed to enhance crop productivity and quality \cite{b13, b14}.
    
    The human interactors (\textit{Remote Users and Mobile Clients}) manage and control all of the Sensors and Actuators (\textit{Irrigation, Drone Imaging/Spraying}) attached to the system. The \textit{IoT Node} incorporates GPS, sensory metadata (i.e., data/type configuration, protocols, data reading/sending frequency) and enables network connectivity with \textit{IoT Field Gateway}. The usage of GPS in precision agriculture allows the analysis of soil property for croplands. And the analyzed results help to form field mapping and decide what type of soil is suitable for a given crop.  The \textit{Mobile Clients} linked to the \textit{IoT Field Gateway} locally with the help of a Smart App and short-distance radio communication technologies such as Bluetooth, Bluetooth Low Energy, ZigBee, etc.
    
    Like remote users, \textit{Mobile Clients} may check the status such as temperature, humidity, pressure, pH, soil nutrient level of the cropland. This status information would be helpful to take any further actions such as irrigation or applying fertilizer. Here, \textit{Fog Node} performs a sort of computation from the edge of the network and sends only required information to the cloud for further processing.
    
    \textit{Cloud} provides different services such as data accumulation, abstraction, integration, remote visualization, analytics, prototype, Smart App development, and more. It also integrates \textit{Satellite Imaging} to capture cropland along with drone imaging for mainly disease forecasting purposes. Therefore, in \textit{Cloud}, every piece of information is stored, processed, and served to remote users.
    
    \subsection{Data Flow Diagram (DFD)}
\begin{figure*} [!h]
	\centerline{\includegraphics[width=.9\textwidth, height=4.5in]{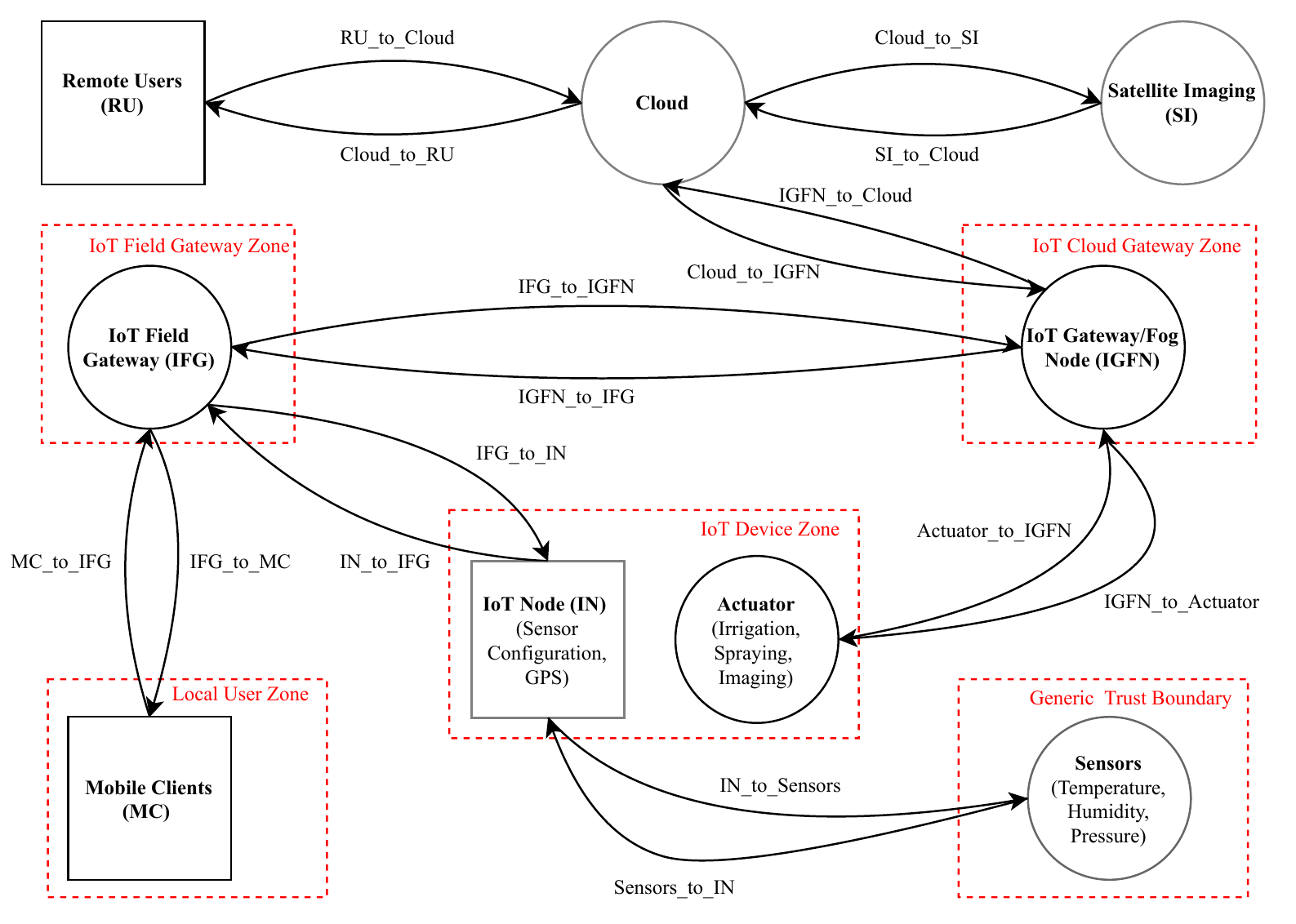}}
 	\caption{Data flow diagram (Out of scope elements: Cloud, Satellite Imaging, IoT Node, and Sensors)}
 	\label{fig-dfd}
	\end{figure*}

    The Fig. \ref{fig-dfd} represents the visual DFD of the precision agriculture system that shows the high-level description and points all of the components along with data flows. Here, we consider the functionalities of all gateways, actuators, and sensors as a process as they accept input data, perform some actions and produce some output information. In the DFD, we have used the circle to denote process, arrow to represent data flow, and rectangle to show external interactor.
    
    \subsection{Threat Modeling}
	Generally, sensor measurements and actuator controlling commands are transmitted through the Internet. So, any compromise may result in a wrong decision or take the control of the overall system. Thus, we cannot deny the potential security issues (especially cyber security) in precision agriculture.
	
	Threat modeling is a proactive way to identify, enumerate, and prioritize threats thus helping to take appropriate safeguards against threats. Simply, it is formed to answer the questions like “Where are the potential threats to the system?”, “What are the most relevant threats?”, and “Where the system is most vulnerable?” \cite{b15}. According to NIST special publication “threat modeling is a form of risk assessment that models aspects of the attack and defense sides of a logical entity, such as a piece of data, an application, a host, a system, or an environment” \cite{b16}.

	\section{EVALUATION OF THREATS}
	In this section, a brief discussion is stated for the evaluation of threats and the tool used to perform the threat modeling.

	\begin{figure*}[!h]
	\centerline{\includegraphics[width=.9\textwidth]{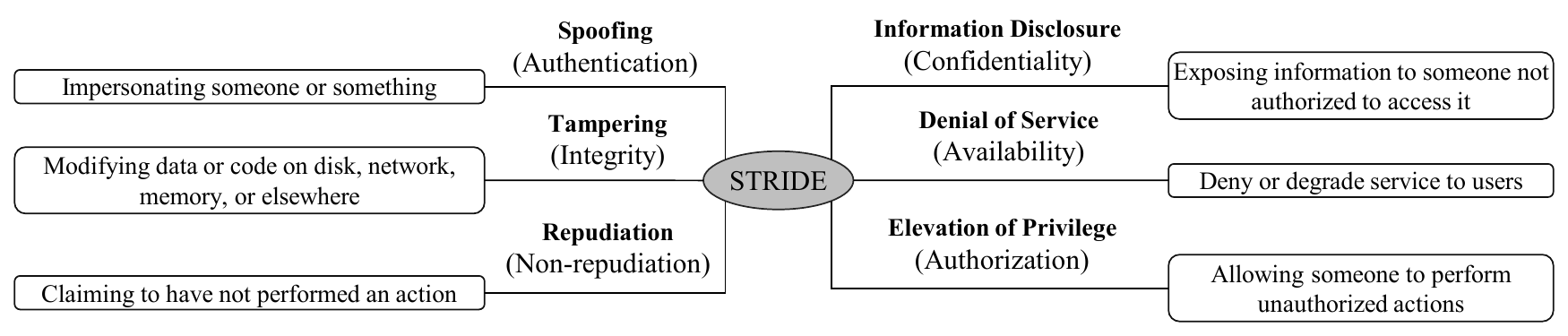}}
 	\caption{Summary of STRIDE threat category with desired security property inside parentheses}
 	\label{fig-stride}
	\end{figure*}

		\begin{table*}[!t]
	 \caption{Identified Threats for selected components, data flows, and interactors}
    \label{table-threatlist}
     \setlength{\tabcolsep}{0.5em} 
	\renewcommand{\arraystretch}{1.3}
	\centering
	\begin{tabularx}{1\textwidth}
	{
      | >{\raggedright\arraybackslash}p{1.45cm} 
      | >{\centering\arraybackslash}p{.85cm} 
      | >{\raggedright\arraybackslash}X |
    }
    \hline
    \textbf{Threat Category} & \textbf{Threat Count}& \textbf{Component, Data Flow, Interactor} \\
    \hline
    \vS & 05 & Remote Users, IoT Gateway/Fog Node, IoT Field Gateway, Actuator, Mobile Clients \\
    \hline
    \vT & 15 & Remote Users, RU_to_Cloud, Cloud_to_RU, IoT Gateway/Fog Node, Cloud_to_IGFN, IGFN_to_Cloud, IoT Field Gateway, IFG_to_IGFN, IGFN_to_IFG, Actuator, Actuator_to_IGFN, IGFN_to_Actuator, Mobile Clients, MC_to_IFG, IFG_to_MC \\
    \hline
    \vR & 05 & Remote Users, IoT Gateway/Fog Node, IoT Field Gateway, Actuator, Mobile Clients \\
    \hline
    \vI & 15 & Remote Users, RU_to_Cloud, Cloud_to_RU, IoT Gateway/Fog Node, Cloud_to_IGFN, IGFN_to_Cloud, IoT Field Gateway, IFG_to_IGFN, IGFN_to_IFG, Actuator, Actuator_to_IGFN, IGFN_to_Actuator, Mobile Clients, MC_to_IFG, IFG_to_MC  \\
    \hline
    \vD & 14 & Remote Users, RU_to_Cloud, Cloud_to_RU, IoT Gateway/Fog Node, Cloud_to_IGFN, IGFN_to_Cloud, IoT Field Gateway, IFG_to_IGFN, IGFN_to_IFG, Actuator, Actuator_to_IGFN, IGFN_to_Actuator, MC_to_IFG, IFG_to_MC \\
    \hline
    \vE & 04 & IoT Gateway/Fog Node, IoT Field Gateway, Actuator, Mobile Clients \\
    \hline
    \end{tabularx}
    \end{table*}
    
    \begin{figure*}[!h]
  	\centerline{\includegraphics[width=.9\textwidth, height=2.25in]{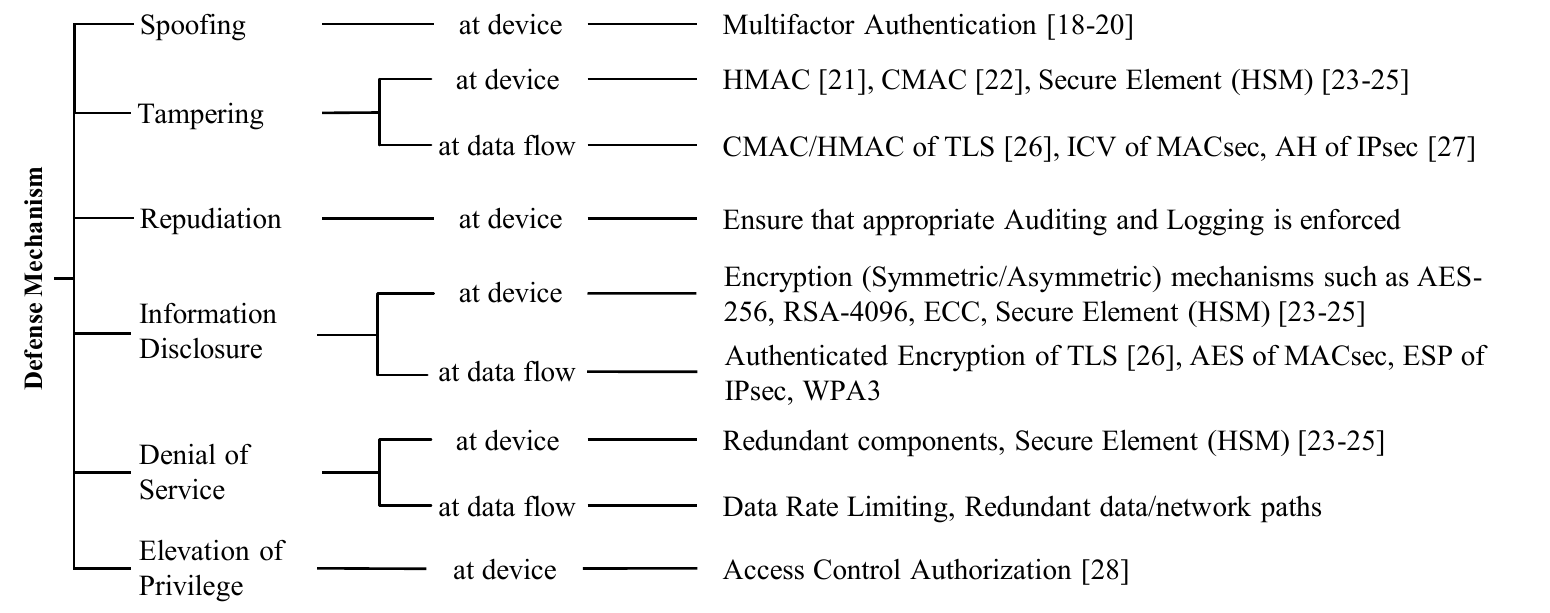}}
 	\caption{Cyber security defense mechanisms against STRIDE category}
 	\label{fig-defense}
	\end{figure*}
	
	\subsection{Analyze Threats in DFD}
	
	In this work, we have performed threat modeling on selected components, data flows, and interactors of DFD. We assume that the Local User Zone, IoT Cloud Zone, IoT Field Gateway Zone, IoT Device Zone, and Local User Zone are physically secure and trusted (which is marked with red dashes in DFD). Besides, we didn’t perform any threat analysis (kept out of scope that marked as blur in DFD) for some components including Cloud, Satellite Imaging, IoT Node, and Sensors. The IoT Node and Sensors are placed in the croplands and trusted thus kept out of scope. Moreover, we believe that the Cloud/Satellite Imaging service providers take appropriate countermeasures for possible security issues as they are accounted for.
	
	Finally, we consider only threats that associated with major components, data flows (interaction), and external interactor. Those are as follows: 1) External interactors: Remote Users (RU), Mobile Clients (MC); 2) Components: IoT Gateway/Fog Node (IGFN), IoT Field Gateway (IFG), Actuator; 3) Data flows: Cloud_to_RU, RU_to_Cloud, IGFN_to_Cloud, Cloud_to_IGFN, IGFN_to_IFG, IFG_to_IGFN, IGFN_to_Actuator, Actuator_to_IGFN, IFG_to_MC, MC_to_IFG.
	
	\subsection{Threat Models}
	There are multiple models available that enable to perform cyber security threats assessment \cite{b17}. For example, Microsoft STRIDE, an acronym that covers six threat categories namely Spoofing, Tampering, Repudiation, Information Disclosure, Denial of Service, and Elevation of Privilege; Process for Attack Simulation and Threat Analysis (PASTA); Common Vulnerability Scoring System (CVSS); Attack trees; Security Cards; Hybrid Threat Modeling Method (hTMM); Trike; Persona non Grata (PnG); LINDDUN (linkability, identifiability, nonrepudiation, detectability, disclosure of information, unawareness, noncompliance); Operationally Critical Threat, Asset, and Vulnerability Evaluation (OCTAVE); Quantitative Threat Modeling Method (Quantitative TMM); Visual, Agile, and Simple Threat (VAST) Modeling. 
	
	Among all, we have chosen the Microsoft STRIDE model to discover threats and possible mitigations because it is widely accepted in industry and academia. Moreover, an open-source (free of cost) tool called “Microsoft Threat Modeling Tool” is available from Microsoft that is continuously updated and handy to perform threat modeling. The summary of the STRIDE model is represented in Fig. \ref{fig-stride}.

	\section{RESULTS AND DISCUSSION}
    After considering components, data flows, interactors, and threat modeling methods, this section exhibited resultant threat details with some defense mechanisms against threat categories.
    
	\subsection{Resultant Threats}
	We have applied the Microsoft STRIDE method (without taking any mitigation actions) on DFD for the chosen components, data flows, and interactors to identify cyber security threats. As a result, we got fifty-eight threats under six STRIDE categories listed in Table-\ref{table-threatlist}.
	
	\subsection{Defense Mechanisms}
	Cyber security defense mechanism aims to protect connected devices (i.e., gateway, router, switch, hub, pc, smartphone) from any compromise or provide software services to authenticated users only. So, the defense is at both hardware and software level that can be the antimalware, firewall, encryption, intrusion detection, access control, authentication, authorization, and more. For securing the data flow and at rest, several defense mechanisms are applied, for example, Hash-based Message Authentication Code (HMAC), Cipher-based MAC (CMAC), Rivest–Shamir–Adleman (RSA) cryptosystem, Advanced Encryption Standard (AES), Elliptic curve cryptography (ECC), Transport Layer Security (TLS), Internet Protocol Security (IPsec), Wi-Fi Protected Access 3 (WPA3), Authentication Header (AH), Encapsulating Security Payload (ESP), Integrity Check Value (ICV), Media Access Control Security (MACsec), and Hardware Security Module (HSM). In this work, we have provided generalized defense mechanisms against STRIDE threat categories presented in Fig. \ref{fig-defense}. Moreover, we suggest following NIST guidelines for standard cryptographic algorithms and key length used to secure data or communication in each threat category \cite{b29,b30}.

	\section{CONCLUSION AND FUTURE WORK}
    Precision agriculture optimizes the farming process, thus increasing productivity by utilizing real-time sensors/devices that help make data-driven decisions. However, cyber security issues evolved from and with these devices and data communication potential stakeholders' concerns. This work has performed threat modeling on a standard precision agriculture framework using the Microsoft STRIDE model to tackle cyber security issues. This work uses the Microsoft STRIDE model to identify, enumerate, and categorize threats for selected components, data flow, and external interactors. As an outcome, there identified fifty-eight cyber security threats which need to be controlled for a thriving smart farming eco-system. As the next stage, this work will lead to deepening each threat to assess risk. So that we can apply the most appropriate mitigation strategy for security betterment in precision agriculture systems is scoped as the future work.
	
	
	


\end{document}